\documentclass[twocolumn,floatfix,showpacs,superscriptaddress]{revtex4}
\pdfoutput=1
\usepackage{graphicx}
\usepackage{amsmath}
\usepackage{bm}
\begin{document}

\title{Theory of the topological Anderson insulator}
\author{C. W. Groth}
\affiliation{Instituut-Lorentz, Universiteit Leiden, P.O. Box 9506, 2300 RA Leiden, The Netherlands}
\author{M. Wimmer}
\affiliation{Instituut-Lorentz, Universiteit Leiden, P.O. Box 9506, 2300 RA Leiden, The Netherlands}
\author{A. R. Akhmerov}
\affiliation{Instituut-Lorentz, Universiteit Leiden, P.O. Box 9506, 2300 RA Leiden, The Netherlands}
\author{J. Tworzyd{\l}o}
\affiliation{Instituut-Lorentz, Universiteit Leiden, P.O. Box 9506, 2300 RA Leiden, The Netherlands}
\affiliation{Institute of Theoretical Physics, Warsaw University, Ho\.{z}a 69, 00--681 Warsaw, Poland}
\author{C. W. J. Beenakker}
\affiliation{Instituut-Lorentz, Universiteit Leiden, P.O. Box 9506, 2300 RA Leiden, The Netherlands}
\date{August 2009}
\begin{abstract}
We present an effective medium theory that explains the disorder-induced transition into a phase of quantized conductance, discovered in computer simulations of HgTe quantum wells. It is the combination of a random potential and quadratic corrections $\propto p^{2}\sigma_{z}$ to the Dirac Hamiltonian that can drive an ordinary band insulator into a topological insulator (having an inverted band gap). We calculate the location of the phase boundary at weak disorder and show that it corresponds to the crossing of a band edge rather than a mobility edge. Our mechanism for the formation of a topological Anderson insulator is generic, and would apply as well to three-dimensional semiconductors with strong spin-orbit coupling.
\end{abstract}
\pacs{73.20.Fz, 03.65.Vf, 73.40.Lq, 73.43.Nq}
\maketitle

Topological insulators continue to surprise with unexpected physical phenomena \cite{But09}. A recent surprise was the discovery of the topological Anderson insulator (TAI) by Li, Chu, Jain, and Shen \cite{Li09}. In computer simulations of a HgTe quantum well, these authors discovered in the phase diagram a transition from an ordinary insulating state (exponentially small conductance) to a state with a quantized conductance of $G_{0}=2e^{2}/h$. The name TAI refers to the latter state. The findings of Ref.\ \cite{Li09} were confirmed by independent simulations \cite{Jia09}. 

The phenomenology of the TAI is similar to that of the quantum spin Hall (QSH) effect, which is well understood \cite{Kan05,Ber06} and observed experimentally in HgTe quantum wells \cite{Koe07,Kon08,Rot09}. The QSH effect is a band structure effect: It requires a quantum well with an inverted band gap, modeled by an effective Dirac Hamiltonian with a negative (socalled ``topological'') mass. The matching of this negative mass inside the system to the usual positive mass outside leaves edge states in the gap. The edge states are ``helical'', in the sense that the direction of propagation is tied to the electron spin. Opposite edges each contribute $e^{2}/h$ to the conductance. The conductance remains quantized in the presence of (weak) disorder, because time reversal symmetry forbids scattering between counter-propagating edge states (of opposite helicity) \cite{Kan05,Ber06,Koe07,Kon08,Rot09}.

The crucial difference between the TAI and QSH phases is that the QSH phase extends down to zero disorder, while the TAI phase has a boundary at a minimal disorder strength. Put differently, the helical edge states in the QSH phase exist \textit{in spite of} disorder, while in the TAI phase they exist \textit{because of} disorder. Note that the familiar quantum Hall effect is like the QSH effect in this respect: The edge states in the quantum Hall effect exist already without disorder (although, unlike the QSH effect, they only form in a strong magnetic field). 

The computer simulations of Refs.\ \cite{Li09,Jia09} confront us, therefore, with a phenomenology without precedent: By what mechanism can disorder produce edge states with a quantized conductance? That is the question we answer in this paper.

We start from the low-energy effective Hamiltonian of a HgTe quantum well, which has the form \cite{Ber06}
\begin{equation}
H=\alpha (p_{x}\sigma_{x}-p_{y}\sigma_{y})+(m+\beta p^{2})\sigma_{z} + [\gamma p^{2}+U(\bm{r})]\sigma_{0}.\label{Hdef}
\end{equation}
This is a two-dimensional Dirac Hamiltonian (with momentum operator $\bm{p}=-i\hbar\nabla$, Pauli matrices $\sigma_{x}$, $\sigma_{y}$, $\sigma_{z}$, and a $2\times 2$ unit matrix $\sigma_{0}$), acting on a pair of spin-orbit coupled degrees of freedom from conduction and valence bands. The complex conjugate $H^{\ast}$ acts on the opposite spin. We assume time reversal symmetry (no magnetic field or magnetic impurities) and neglect any coupling between the two spin blocks $H$ and $H^{\ast}$ \cite{note2}. The scalar potential $U$ accounts for the disorder. The parameters $\alpha,\beta,\gamma,m$ depend on the thickness and composition of the quantum well \cite{Kon08}. For the specific calculations that follow, we will use the same parameter values as in Ref.\ \cite{Li09}, representative of a non-inverted HgTe/CdTe quantum well \cite{param}.

The terms quadratic in momentum in Eq.\ \eqref{Hdef} are not present in the Dirac Hamiltonian familiar from relativistic quantum mechanics, but they play an important role here. In particular, it is the relative sign of $\beta$ and $m$ that determines whether the clean quantum well ($U\equiv 0$) is inverted ($\beta m<0$) or not-inverted ($\beta m>0$). We take $\beta>0$, so the inverted quantum well has a negative topological mass $m<0$. The inverted quantum well is a topological insulator (for Fermi energies $E_{F}$ inside the gap), while the non-inverted quantum well is an ordinary band insulator. The phase transition between these two types of insulators therefore occurs at $m=0$ in a clean quantum well.

We will now show that disorder can push the phase transition to positive values of $m$, which is the hallmark of a TAI. Qualitatively, the mechanism is as follows. Elastic scattering by a disorder potential causes states of definite momentum to decay exponentially as a function of space and time. The quadratic term $\beta p^{2}=-\hbar^{2}\beta\nabla^{2}$ in $H$, acting on the decaying state $\propto e^{-x/\lambda}$, adds a negative correction $\delta m$ to the topological mass. The renormalized mass $\bar{m}=m+\delta m$ can therefore have the opposite sign as the bare mass $m$. Topological mass renormalization by disorder, and the resulting change in the phase diagram, has previously been studied without the terms quadratic in momentum \cite{Shi09}. The sign of $\bar{m}$ and $m$ then remains the same and the TAI phase cannot appear.

We extract the renormalized topological mass $\bar{m}$, as well as the renormalized chemical potential $\bar{\mu}$, from the self-energy $\Sigma$ of the disorder-averaged effective medium. To make contact with the computer simulations \cite{Li09,Jia09}, we discretize $H$ on a square lattice (lattice constant $a$) and take a random on-site disorder potential $U$, uniformly distributed in the interval $(-U_{0}/2,U_{0}/2)$. We denote by $H_{0}(\bm{k})$ the lattice Hamiltonian of the clean quantum well in momentum representation \cite{Ber06,Kon08}. 

The self-energy, defined by
\begin{equation}
(E_{F}-H_{0}-\Sigma)^{-1}=\langle(E_{F}-H)^{-1}\rangle, \label{sigmadef}
\end{equation}
with $\langle\cdots\rangle$ the disorder average, is a $2\times 2$ matrix which we decompose into Pauli matrices: $\Sigma=\Sigma_{0}\sigma_{0}+\Sigma_{x}\sigma_{x}+\Sigma_{y}\sigma_{y} +\Sigma_{z}\sigma_{z}$. The renormalized topological mass and chemical potential are then given by
\begin{equation}
\bar{m}=m+\lim_{k\rightarrow 0}{\rm Re}\,\Sigma_{z},\;\;\bar{\mu}=E_{F}-\lim_{k\rightarrow 0}{\rm Re}\,\Sigma_{0}.\label{barmbarmu}
\end{equation}
The phase boundary of the topological insulator is at $\bar{m}=0$, while the Fermi level enters the (negative) band gap when $|\bar{\mu}|=-\bar{m}$. 

\begin{figure}[tb]
\centerline{\includegraphics[width=1\linewidth]{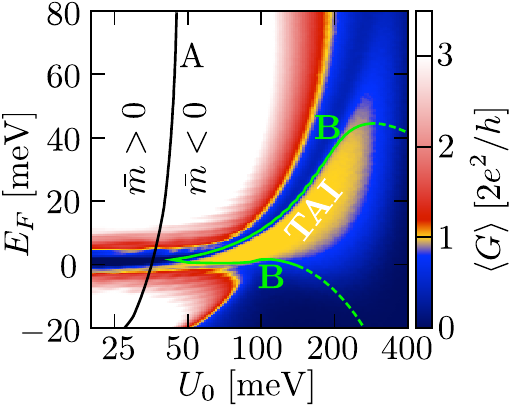}}
\caption{\label{fig_phasediagram}
Computer simulation of a HgTe quantum well \cite{param}, showing the average conductance $\langle G\rangle$ as a function of disorder strength $U_{0}$ (logarithmic scale) and Fermi energy $E_{F}$, in a disordered strip of width $W=100\,a$ and length $L=400\,a$. The TAI phase is indicated. Curves A and B are the phase boundaries resulting from the effective medium theory. Curve A separates regions with positive and negative renormalized topological mass $\bar{m}$, while curve B marks the crossing of the renormalized chemical potential $\bar{\mu}$ with the band edge ($|\bar{\mu}|=-\bar{m}$). Both curves have been calculated {\em without any adjustable parameter}. The phase boundary of the TAI at strong disorder is outside of the regime of validity of the effective medium theory. 
}
\end{figure}

In the selfconsistent Born approximation, $\Sigma$ is given by the integral equation \cite{She06}
\begin{equation}
\Sigma=\tfrac{1}{12}U_{0}^{2}(a/2\pi)^{2}\int_{\rm BZ} d\bm{k}\, [E_{F}+i0^{+}-H_{0}(\bm{k})-\Sigma]^{-1}.\label{eq:scba}
\end{equation}
(The integral is over the first Brillouin zone.) The self-energy is independent of momentum and diagonal (so there is no renormalization of the parameters $\alpha,\beta,\gamma$). By calculating $\bar{m}$ and $\bar{\mu}$ as a function of $E_{F}$ and $U_{0}$ we obtain the two curves A and B in Fig.\ \ref{fig_phasediagram}.

We have also derived an approximate solution in closed form \cite{note1},
\begin{subequations}\label{closedform}
\begin{align}
&\bar{m}=m-\frac{U_{0}^{2}a^{2}}{48\pi\hbar^{2}} \frac{\beta}{\beta^2-\gamma^2} \ln\left|\frac{\beta^2-\gamma^2}{E_F^2-m^2}\left(\frac{\pi\hbar}{a}\right)^{4} \right|,\label{closedformm}\\
&\bar{\mu}=E_{F}-\frac{U_{0}^{2}a^{2}}{48\pi\hbar^{2}} \frac{\gamma}{\beta^2-\gamma^2} \ln\left|\frac{\beta^2-\gamma^2}{E_F^2-m^2}\left(\frac{\pi\hbar}{a}\right)^{4} \right|,\label{closedformmu}
\end{align}
\end{subequations}
showing that the correction $\delta m=\bar{m}-m$ to the topological mass by disorder is {\em negative\/} --- provided $\beta>\gamma$. For $\beta<\gamma$ the clean HgTe quantum well would be a semimetal, lacking a gap in the entire Brillouin zone. Neither the TAI phase nor the QSH phase would then appear. In HgTe the parameter $\beta$ is indeed larger than $\gamma$, but not by much \cite{param}. Eq.\ \eqref{closedform} implies that the lower branch of curve B (defined by $\bar{\mu}=\bar{m}<0$) is then fixed at $E_{F}\approx m>0$ independent of $U_{0}$. This explains the puzzling absence of the TAI phase in the valence band ($E_{F}<0$), observed in the computer simulations \cite{Li09,Jia09}.

\begin{figure}[tb]
\centerline{\includegraphics[width=0.8\linewidth]{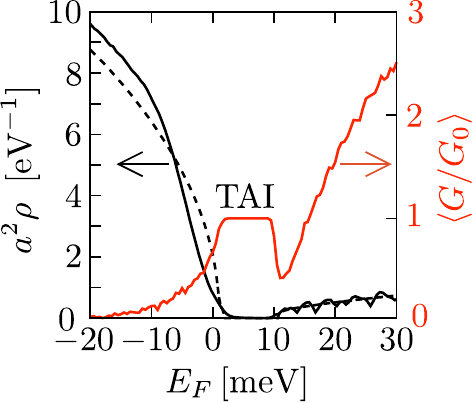}}
\caption{\label{fig_dos}
Black curves, left axis: Average density of states $\rho$ as a function of $E_{F}$ for $U_{0}=100\,{\rm meV}$, calculated by computer simulation (solid curve, for a disordered $100\,a\times 100\,a$ square with periodic boundary conditions) or by effective medium theory (dashed curve). Red curve, right axis: Average conductance $\langle G\rangle$, calculated by computer simulation in a disordered strip  ($U_{0}=100\,{\rm meV}$, $W=100\,a$, $L=400\,a$). The TAI phase of quantized conductance lines up with the band gap.
}
\end{figure}

To quantitatively test the phase diagram resulting from the effective medium theory, we performed computer simulations similar to those reported in Refs.\ \cite{Li09,Jia09}. The conductance $G$ is calculated from the lattice Hamiltonian \cite{Ber06,Kon08} in a strip geometry, using the method of recursive Green functions. The strip consists of a rectangular disordered region (width $W$, length $L$), connected to semi-infinite, heavily doped, clean leads \cite{note4}. Theory and simulation are compared in Figs.\ \ref{fig_phasediagram} and \ref{fig_dos}. 

Fig.\ \ref{fig_phasediagram} shows the phase diagram. The weak-disorder boundary of the TAI phase observed in the simulations is described quite well by the selfconsistent Born approximation (curve B) --- {\em without any adjustable parameter}. Curve B limits the region where (A) the renormalized topological mass $\bar{m}$ is negative {\em and\/} (B) the renormalized chemical potential $\bar{\mu}$ lies inside the band gap: $|\bar{\mu}|<-\bar{m}$. Condition (A) is needed for the existence of edge states with a quantized conductance. Condition (B) is not needed for an infinite system, because then Anderson localization suppresses conductance via bulk states as well as coupling of edge states at opposite edges. In the relatively small systems accessible by computer simulation, the localization length for weak disorder remains larger than the system size (see later). Condition (B) is then needed to eliminate the bulk conductance and to decouple the edge states.

Fig.\ \ref{fig_dos} shows the average density of states $\rho$ at the Fermi level. The agreement between the selfconsistent Born approximation (dashed black curve) and the computer simulation (solid black) is quite good, in particular considering the fact that this plot is for a disorder strength which is an order of magnitude larger than the band gap. The range of Fermi energies over which the gap extends lines up nicely with the  conductance plateau, shown in the same figure (red curve).

The strong-disorder phase boundary of the TAI cannot be described by effective medium theory, but it should be similar to the QSH phase boundary. In the QSH effect the strong-disorder transition is in the universality class of the quantum Hall effect \cite{Ono03} --- in the absence of coupling between the spin blocks \cite{Ono07,Obu07}. To ascertain the nature of the strong-disorder transition out of the TAI, we have calculated the critical exponent $\nu$ governing the scaling of the localization length $\xi$. For that purpose we roll up the strip into a cylinder, thereby eliminating the edge states \cite{Jia09}. We determine the localization length $\xi\equiv-2\lim_{L\rightarrow\infty}L\langle\ln G/G_{0}\rangle^{-1}$ by increasing the length $L$ of the cylinder at fixed circumference $W$.

\begin{figure}[tb]
\centerline{\includegraphics[width=0.7\linewidth]{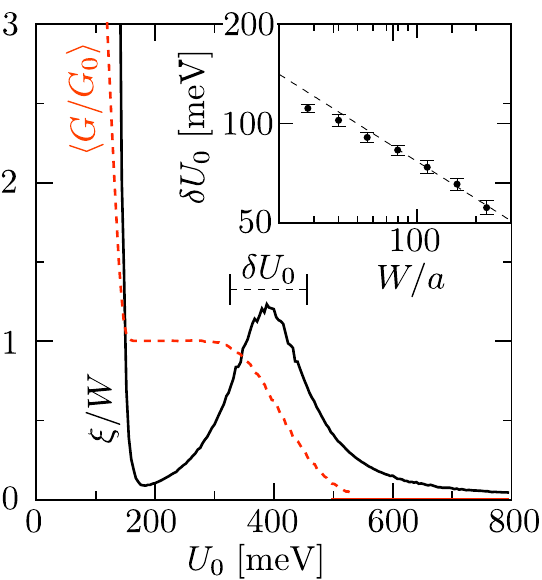}}
\caption{\label{fig_scaling}
Red dashed curve: Average conductance $\langle G\rangle$ as a function of disorder strength ($E_{F}=25\,{\rm meV}$, $W=L=100\,a$). Black solid curve: Localization length $\xi$, showing the peak at the strong-disorder edge of the conductance plateau --- characteristic of a localization transition. The scaling with system size $W$ of the width $\delta U_{0}$ of the peak is shown in the inset (double-logarithmic plot).
}
\end{figure}

In Fig.\ \ref{fig_scaling} we show $\xi$ as a function of disorder strength $U_{0}$ at $E_{F}=25\,{\rm meV}$, $W=100\,a$. As mentioned above, $\xi$ becomes much larger than $W$ upon crossing the weak-disorder boundary of the TAI, so there is no localization there \cite{note7}. At the strong-disorder boundary, however, the dependence of $\xi$ on $U_{0}$ shows the characteristic peak of a localization transition \cite{Eve08}. In the inset we plot the scaling with $W$ of the width $\delta U_{0}$ at half maximum of the peak. This yields the critical exponent via $\delta U_{0}\propto W^{-1/\nu}$. We find $\nu=2.66\pm 0.15$, consistent with the value $\nu=2.59$ expected for a phase transition in the quantum Hall effect universality class \cite{Sle09}.

In conclusion, we have identified the mechanism for the appearance of a disorder-induced phase of quantized conductance in computer simulations of a HgTe quantum well \cite{Li09,Jia09}. The combination of a random potential and quadratic momentum terms in the Dirac Hamiltonian can change the sign of the topological mass, thereby transforming a non-inverted quantum well (without edge states in the band gap) into an inverted quantum well (with edge states). The weak-disorder boundary in the phase diagram of the TAI has been calculated by effective medium theory, in good agreement with the simulations (curve B in Fig.\ \ref{fig_phasediagram}). 

Contrary to what the name ``topological Anderson insulator'' might suggest, we have found that the hallmark of the TAI in the simulations, the weak-disorder transition into a phase of quantized conductance, is not an Anderson transition at all. Instead, the weak-disorder boundary B marks the crossing of a band edge rather than a mobility edge. A mobility edge (similar to the QSH effect \cite{Ono03,Ono07,Obu07}) is crossed at strong disorder, as evidenced by the localization length scaling (Fig.\ \ref{fig_scaling}).

Our findings can be summed up in one sentence: ``A topological insulator {\em wants\/} to be topological''. The mechanism for the conversion of an ordinary insulator into a topological insulator that we have discovered is generically applicable to narrow-band semiconductors with strong spin-orbit coupling (since these are described by a Dirac equation, which generically has quadratic momentum terms \cite{Zha09}). There is no restriction to dimensionality. We expect, therefore, a significant extension of the class of known topological insulators \cite{But09} to disordered materials without intrinsic band inversion.

We have benefited from discussions with F. Guinea, A. D. Mirlin, P. M. Ostrovsky, and M. Titov. This research was supported by the Dutch Science Foundation NWO/FOM and by an ERC Advanced Investigator Grant.


\begin{thebibliography}{99}
\bibitem{note2} We have repeated the calculations of the conductance including a coupling Hamiltonian between the spin blocks of the form $\pm i\kappa\sigma_{y}$, with $\kappa=1.6\,{\rm meV}$, representative of bulk inversion asymmetry in a HgTe quantum well. The effect on the phase diagram was negliglibly small.
\bibitem{param} The parameter values of $H$ that we have used are: $\hbar\alpha=364.5\,{\rm meV}\,{\rm nm}$, $\hbar^{2}\beta=686\,{\rm meV}\,{\rm nm}^{2}$, $\hbar^{2}\gamma=512\,{\rm meV}\,{\rm nm}^{2}$, $m=1\,{\rm meV}$. The lattice constant of the discretization was $a=5\,{\rm nm}$.
\bibitem{note4} The reason we dope the leads in the computer simulations is to be able to access the region $|E_{F}|<m$ in the phase diagram, where the band gap in the clean leads would otherwise prevent conduction through the disordered region.
\bibitem{note7} A pronounced conductance dip between the quantized plateau and the high-conductance regime exists for samples with a large aspect ratio $L/W$, becoming broader and broader with increasing $L/W$. The dip is clearly visible in Fig.\ \ref{fig_dos} (for $L/W=4$) and absent in Fig.\ \ref{fig_scaling} (for $L/W=1$). Our numerical data suggests that the conductance dip extends over the parameter range where $W<\xi<L$, so that conduction is suppressed both through the bulk and along the edges.
\end{thebibliography}
\end{document}